\documentclass[twocolumn]{WileyMSP-template}
\usepackage{microtype}
\usepackage{float}
\usepackage{ragged2e}
\usepackage{subcaption}
\usepackage[utf8]{inputenc}
\usepackage{placeins}
\usepackage{setspace}
\usepackage{changes}
\usepackage[margin=0.5in]{geometry}

\begin{document}

\UseRawInputEncoding
\justifying

\pagestyle{fancy}
\rhead{}

\twocolumn[
\begin{@twocolumnfalse}
\title{Formation of Light-Emitting Defects
in Ag-based Memristors}

\maketitle

\vspace{0.6 cm}


\author{Diana Singh}
\author{Maciej \'Cwierzona}
\author{R\'egis Parvaud}
\author{Sebastian Ma\'ckowski} 
\author{Alexandre Bouhelier*}


\vspace{0.35 cm}


\begin{affiliations}
Diana Singh, Maciej \'Cwierzona, R\'egis Parvaud, Alexandre Bouhelier\\
Universit\'e Bourgogne Europe, CNRS, Laboratoire Interdisciplinaire Carnot de Bourgogne ICB UMR 6303, 21000 Dijon, France\\
Email: diana.singh@u-bourgogne.fr, regis.parvaud@u-bourgogne.fr, alexandre.bouhelier@u-bourgogne.fr

Maciej \'Cwierzona, Sebastian Ma\'ckowski\\
Nicolaus Copernicus University, Institute of Physics, Faculty of Physics, Astronomy and Informatics, 87-100 Toru\'n, Poland\\
Email: mcwierzona@umk.pl, mackowski@fizyka.umk.pl

Alexandre Bouhelier\\
Universit\'e de Sherbrooke, CNRS, Laboratoire International Fronti\`eres Quantiques, IRL2017, Sherbrooke, QC, Canada

ORCID: Singh Diana (0009-0007-3063-9138); Maciej \'Cwierzona (0000-0003-4365-6440);\\ Sebastian Ma\'ckowski (0000-0003-1560-6315); Alexandre Bouhelier (0000-0002-0391-2836).

\end{affiliations}

\keywords{Memristors, Resistive switching, Electroluminescence, Photoluminescence, Ag cluster, Metal filament}

\begin{abstract}

Optical memristors are innovative devices that enable the integration of electro-optical functionalities—such as light modulation, multilevel optical memory, and nonvolatile reprogramming—into neuromorphic networks. Recently, their capabilities have expanded with the development of light-emitting memristors, which operate through various emission mechanisms. One notable process involves the electroluminescence of defects generated within the switching matrix during device activation. In this study, we explore the early-stage formation and evolution of the species responsible for light emission in Ag-based in-plane memristors. Our approach combines electrical stimulation with correlated optical electroluminescence and photoluminescence measurements. The findings provide valuable insights into controlling emission processes in memristors, paving the way for their integration as essential components in neuromorphic circuits.

\end{abstract}
\vspace{1 cm}
\end{@twocolumnfalse}
]
\section{Introduction}

Driven by the surging global demand for data processing, researchers in both academic and industrial settings are actively exploring new device architectures beyond traditional transistor-based systems~\cite{caulfield2010, mehonic2024}. Among the most promising alternatives is memristive electronics~\cite{Lanza2025, alharbi2024}. This technology harnesses the compactness of a two-terminal nonlinear device~\cite{Kim-ACSNano23} to enable energy-efficient, high-performance circuits ideal for advanced applications like in-memory computing~\cite{Ielmini2018,LeGallo2023, qin2025} and neuromorphic networks capable of emulating complex cognitive functions ~\cite{Wang22,Weilenmann2024}.
Beyond purely electronic solutions, the field is expanding into hybrid circuits that integrate photonic components~\cite{Atabaki2018, mao2019} -- including quantum systems~\cite{Kramnik2025} -- to accelerate data transfer, minimize heat dissipation, and support quantum infrastructure. Within this rapidly evolving optoelectronic landscape, several teams have successfully integrated photonic analogs to memristors~\cite{DiMartino2019,Youngblood2023}. Like their electronic counterparts, these new devices can dynamically modulate light intensity in waveguide-based systems~\cite{Singh2021} or act as optically and electrically-read switches~\cite{Chen:14,Farmakidis2019}. These features are key in the development optoelectronic memristors~\cite{Pereira2023} used for instance in photomemristive reservoir networks~\cite{Tan2023}. A standout feature of memristive systems is their capacity to emit radiation during resistive switching~\cite{Zakhidov2010} effectively combining optical and electronic functions within a single nanoscale device~\cite{Cheng2022,Zhou2026}. Light-emitting memristors are designed using various approaches. Some architectures incorporate active materials within the memristor, like quantum dots~\cite{Li21,Yen2021,Zhu2022} or two-dimensional compounds~\cite{He2013}. While encouraging, these routes call from further material development and device engineering to achieve the required stability and uniformity. Other device architectures self-generate the species responsible for electroluminescence ~\cite{Cheng2022,Malchow:21,Hamdad2023} and are the focus of this report. Here we investigate the formation dynamics of native optically active species spontaneously generated during the activation phase of  memristive devices. Namely, we characterize the conditions leading to the onset of electroluminescence in planar Ag-based memristors.
The experimental approach, combining electrical stimulation with \emph{in-situ} photoluminescence measurements, enables us to monitor the time evolution of light-emitting species during device's activation. We detect a strongly fluctuating photoluminescence signal from the switching matrix, driven by the injection and diffusion of metal precursors. Our observations show that photoluminescence serves as a distinctive probe for capturing the early-stage dynamics that lead to the spontaneous formation of optically active species within the memristive structure. Our results offer deeper insights into the underlying diffusive mechanisms at play and contribute to providing a general understanding of light-emitting memristors.

\section{Nanofabrication and Experimental Setup}
The memristor design used in this study consists of two tapered Ag electrodes, separated by a gap of approximately 300 nm, deposited on a standard glass coverslip. A Poly(methyl methacrylate) (PMMA) polymer cover layer acts as a dielectric switching matrix and protects the silver from oxidation. 

The rationale behind this device configuration is the following. First, with Ag-based memristors, the change in conductance is mediated by the electrochemical growth of a metal filament in the switching matrix during the so-called electrical activation phase~\cite{Wei12,Cheng_19}. Electroluminescence (EL), when present, is observed only after concluding this initial process ~\cite{Cheng2022,Malchow2024}. This suggest the occurrence of a dynamical process preceding the emission, which is linked to the filament formation. This is precisely what we are interested in elucidating and understanding. Second, to enable optical resolution of features forming within the inter-electrode gap, we intentionally select a gap size on the order of the microscope’s spatial resolution. This geometry, however, entails a trade-off in overall memristive performance relative to more advanced technologies. Consequently, key metrics such as switching speed, endurance, and retention time fall short of those achieved in best-in-class architectures.
Third, we employ a polymer host rather than conventional oxide-based matrices. This material is not only low-cost and easy to process~\cite{Tsuruoka2024} — making it particularly suitable for planar device configurations — but also facilitates the diffusion of conductive species within the gap~\cite{Jinesh16,Wolf_2015} and helps protect silver from environmental degradation. Naturally, the use of a polymer also contributes to suboptimal electrical performance compared to optimized dielectric matrices~\cite{Sung2022,Tsuruoka2024,Sun2025,Passerini2025}; however, it facilitates our goal of probing the early stages leading to the formation of light-emitting species. Finally and importantly, devices using PMMA as the switching layer have been shown to exhibit light emission~\cite{Cheng2022,Malchow2024}, although previous studies have not offered the level of insight provided here.

\begin{figure}[!ht]
    \centering
        \includegraphics[width=\columnwidth]{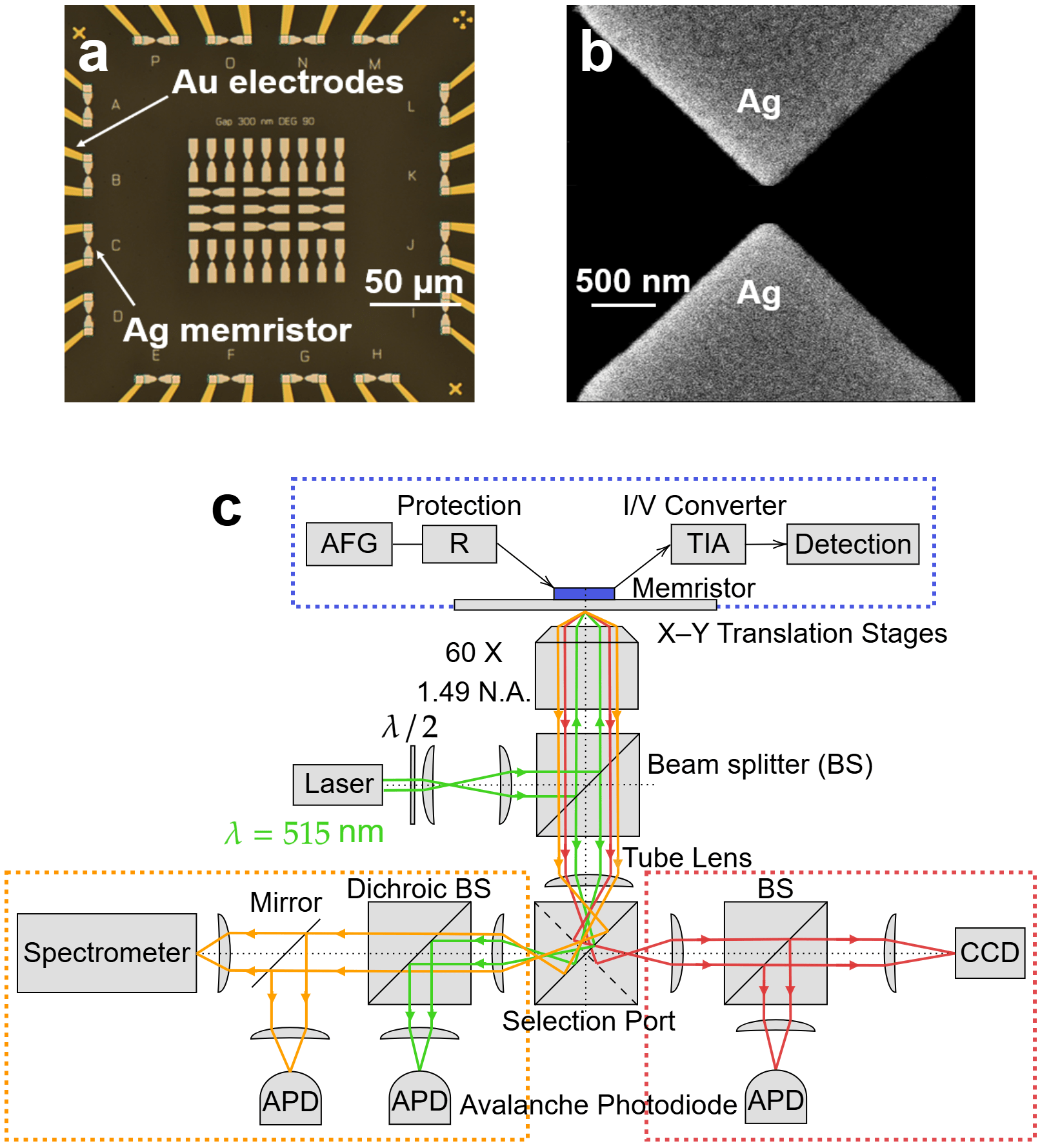}
    \caption{\textbf{a} Bright-field optical image of the sample. It contains 16 connected memristors (labeled A to P), and a series of isolated structures at the center for characterization purpose. The entire coverslip is covered by a 160 nm-thick Poly(methyl methacrylate) layer. \textbf{b} Scanning electron micrograph of a typical memristive gap. It consists of the two Ag tapered electrodes separated by a gap of 300 nm. \textbf{c} Experimental setup. The activation of the memristor and repetitive resistive switching are induced by applying voltage pulses from an arbitrary function generator. The current flowing is measured using a trans-impedance amplifier acting as a current-to-voltage converter. A protection resistance $R$ limits the current. The photoluminescence (PL) is excited with a 515 nm continuous laser tightly focused to a diffraction limited spot on the device. The same objective collects the PL emission (shown in orange) and the laser reflection (shown in green). The sample is raster scanned and these signals are detected by avalanche photodiodes at each positions to reconstruct spatial maps. PL spectra are obtained by a spectrometer. Finally, electroluminescence (shown in red) is detected by a third avalanche photodiode and a charge-coupled device. }
\label{Fig:setup}
\end{figure}

The memristors are fabricated by conducting several technological steps. First, ultraviolet lithography is performed to define a series of macroelectrodes enabling electrical connection to the electrical instrumentation. This step uses a MJB4 Mask Aligner SUSS at a wavelength of 345 nm, and a power of about 15 mW$\cdot$cm$^{-2}$. Macroelectrodes (a set of two per memristor) are made of Au (thickness 70 nm) on top of a Ti adhesion layer (thickness 10 nm). Second, electron-beam lithography (Raith E-line Plus writer operated at 20 kV acceleration voltage and a current of about 20 pA) is conducted to define secondary electrodes connected to the macroelectrodes . They consist of two tapers with an opening angle of 90$^{\circ}$ facing each other and separated by a 300 nm gap. A 2 nm-thick Ti layer followed by a 50 nm-thick Ag layer is then evaporated on the patterned substrate prior to lifting off the electrosensitive resist. Each Au macroelectrode is thus in electrical contact with the corresponding secondary Ag layer. All metallic layers are evaporated using a Plassys MEB400 operated at a pressure below 10$^{-7}$ mbar. The Au and Ag are deposited at 1 nm$\cdot$s$^{-1}$, and the Ti layer at 0.3 nm$\cdot$s$^{-1}$. Finally, a PMMA layer with a thickness of 160 nm is spin-coated across the entire surface of the glass coverslip with the help of Primus STT15 coater. An optical microscope image of a sample and a typical scanning electron micrograph (SEM) of the gap area are shown in Figures~\ref{Fig:setup}a and \ref{Fig:setup}b, respectively. Here, the coverslip was coated with a conductive resist to enable electron imaging.  

The glass substrate supporting the devices is placed on an inverted optical microscope (Ti-U, Nikon) which works as a sample-scanning confocal microscope. The stand is fitted with three distinct modules, as illustrated in Figure~\ref{Fig:setup}c, with the purpose of electrically stimulating the memristor and monitoring in real-time the electrical and optical changes accompanying the formation of Ag filaments. The upper space of the sketch represents the electrical stimulation (blue dotted box). An arbitrary function generator (AFG-3102, Tektronix) delivers $n$ sequences of $N$ voltage pulses of amplitude $V$, period $T$, and duty cycle $D$ defined as the ratio between the pulse duration to the period. A series resistance $R$ protects the device by limiting the current. The current is measured using a transimpedance amplifier (TIA, DHCPA-100, Femto-GmbH). The voltage train and the current reading are both digitized by an acquisition board (Moku:Pro, Liquid Instruments). 
The other two modules are dedicated to optical characterizations. The right exit port of the microscope records the electroluminescence (EL) activity (red dotted box). The EL signal emitted by the memristor is collected by a high numerical aperture objective (N.A. = 1.49) and is directed to an Avalanche Photodiode (APD, SPCM-AQRH, Perkin Elmer) and a Charge-Coupled Device (CCD, iKon, Andor). The APD records the time evolution of the EL, while the CCD provides a time-integrated image of the EL response. The left port of the confocal microscope is used for exciting and detecting the photoluminescence (PL) produced by the device (orange dotted box). A 515 nm laser excitation (Becker \& Hickl GmbH, BDL-515-SMC) is focused by the objective to a diffraction-limited area on the device. The PL and the intensity of the laser reflected back from the sample are spectrally separated by a dichroic beam splitter and detected simultaneously by a pair of APDs to form two-dimensional confocal PL and reflection images when the sample is scanned through the focus. Finally, a grating spectrometer (Shamrock, Andor) measures the spectral content of the PL.

\section{Electrical activation of the device}

\begin{figure}[!ht]
    \centering
        \includegraphics[width=\columnwidth]{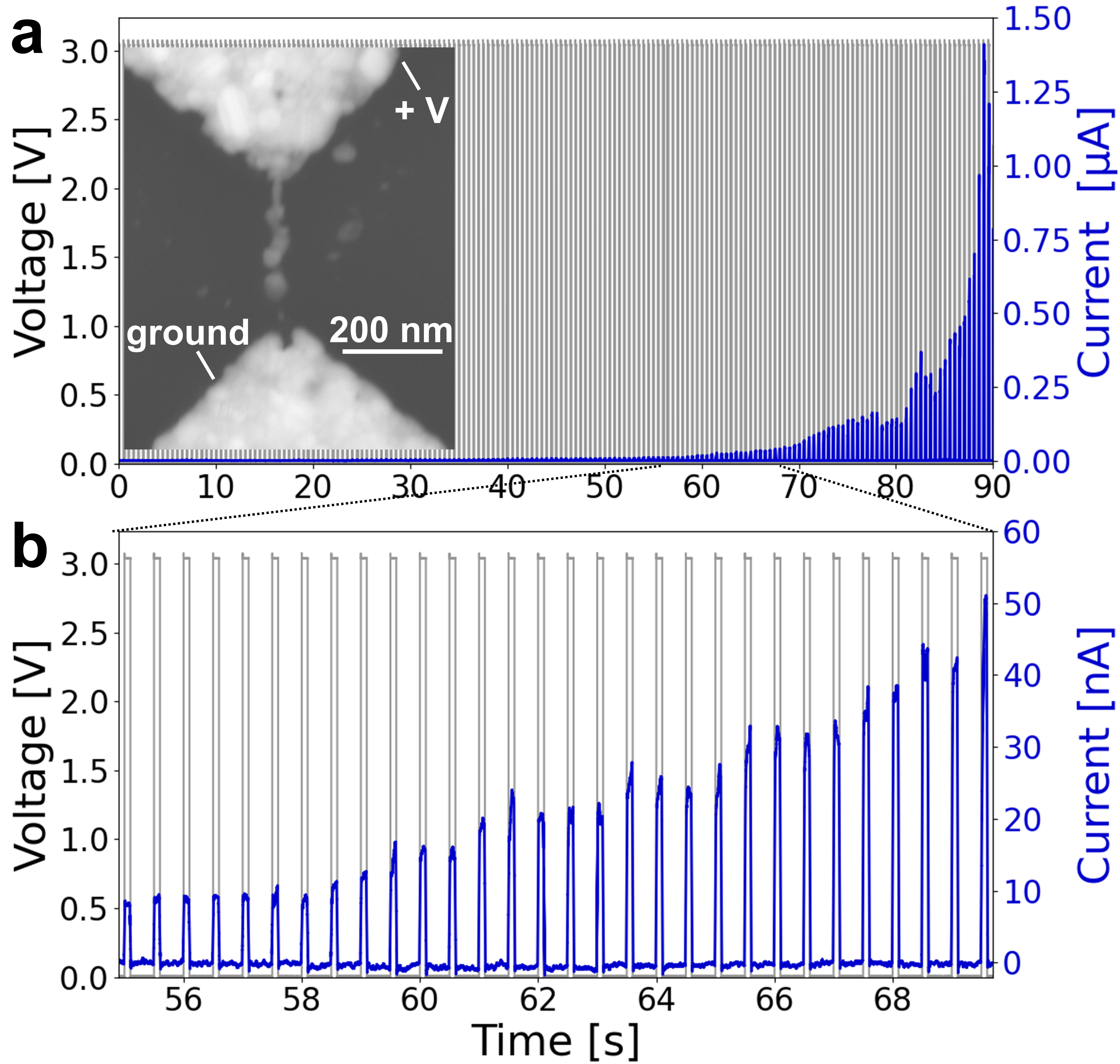}
\caption{\textbf{a} Time plot of the voltage and the current at the end of the activation phase. $V=3\ \mathrm{V}$, $T=500\ \mathrm{ms}$, and $D=0.2$. Inset: SEM image revealing a filamentary structure between the two Ag electrodes formed during the activation of the device. \textbf{b} Time extract showing the rise of the current (\emph{i.e.} the potentiation) during 30 consecutive pulses.}

 \label{fig:activation}
\end{figure}

Before the activation, the 300 nm gap separating the two electrodes prevents any current from flowing. The purpose of the activation phase is to promote the buildup of conduction channels in the dielectric gap. This process is well understood and involves the field-assisted diffusion of Ag ions formed by a redox reaction, their nucleation, and eventually the growth of a conductive filament that will be responsible for resistive switching~\cite{Yang2016,Dirkmann2023}. The change of the device’s conduction, signaling the successful completion of the activation process, is materialized by the flow of current. An example is illustrated in Figure~\ref{fig:activation}a where we trace the current evolution during the last of the $n=49$ activation sequences, which consist of electrical pulses of amplitude $V=3$~V, period $T=500$~ms, and duty ratio $D=0.2$. The protection resistance is here $R=6.12$~k$\Omega$. 
The excitation parameters and the current-limiting resistance are chosen to maintain the device in a volatile state between pulses~\cite{Wang2019}. With this configuration, the conductive pathway relaxes in less than 1 ms (data not shown, but in line with previous works~\cite{Wang2017,Loh2025}), which is significantly shorter than the 400 ms interval between voltage pulses, thereby ensuring volatility for a prolonged measurement time. For Ag-based memristor, structurally unstable filaments are dewetting into a number of disconnected clusters~\cite{Hsiung2010} and their retention time scales with their thickness~\cite{Wang2019}. With these operating conditions, we favor filament dynamics over optimizing device performance. As a result, a larger number of pulse sequences is required to observe the onset of conduction. Although increasing $V$ or 
$D$ accelerates activation, such conditions are not suitable for the experiments presented later in this work.

About $N=120$ pulses in this final sequence are necessary before a measurable current appears. This accumulative process reflects the formation kinetics of the conductive path between the electrodes. Once the resistive switching is established, current is detected at each subsequent voltage pulse with increasing amplitude as the filament grows thicker. A zoom-in view of the so-called potentiation phase is displayed in Figure~\ref{fig:activation}b. 

The repeatability of the resistive switch across devices is difficult to scale to yield large statistical samples. The activation protocol was designed and optimized to maximize filament dynamics after the activation phase. It requires prolonged stimulation because, with the duty ratio used in the experiment, the device relaxes between each pulse. Nonetheless, to demonstrate the device-to-device variability of the forming phase~\cite{roldan2023}, we performed additional experiments in which we stimulated 12 new devices bearing similar geometrical characteristics with a ramping activation voltage instead of a pulsed stimulation. All devices exhibit similar activation characteristics, with an average switching voltage (defined as the voltage at which the current rises above 100 pA) of 21$\pm$1.21V. 
The inset of Figure~\ref{fig:activation}a shows a SEM of the grown filament taken after the activation. For this image, the PMMA layer was removed in acetone prior to sputtering a conductive carbon coating. The image reveals a filament bridging the two electrodes and extending along the SiO$_2$ interface, which is consistent with the expectation that Ag ion nucleation and diffusion is confined within the electrode thickness (ca. 50 nm).

Unfortunately, we are unable to provide electron microscopy images of the activated gaps before and after removal of the PMMA switching layer. As PMMA is an electron-sensitive resist, high-resolution imaging induces irreversible hardening of the polymer, preventing subsequent liftoff using standard Ag-compatible solvents. Consequently, we cannot rule out the presence of additional filamentary pathways forming through the thickness of the PMMA, which may be removed along with the resist during the liftoff process prior to imaging.

\begin{figure}[htbp]
    \centering
        \includegraphics[width=\columnwidth]{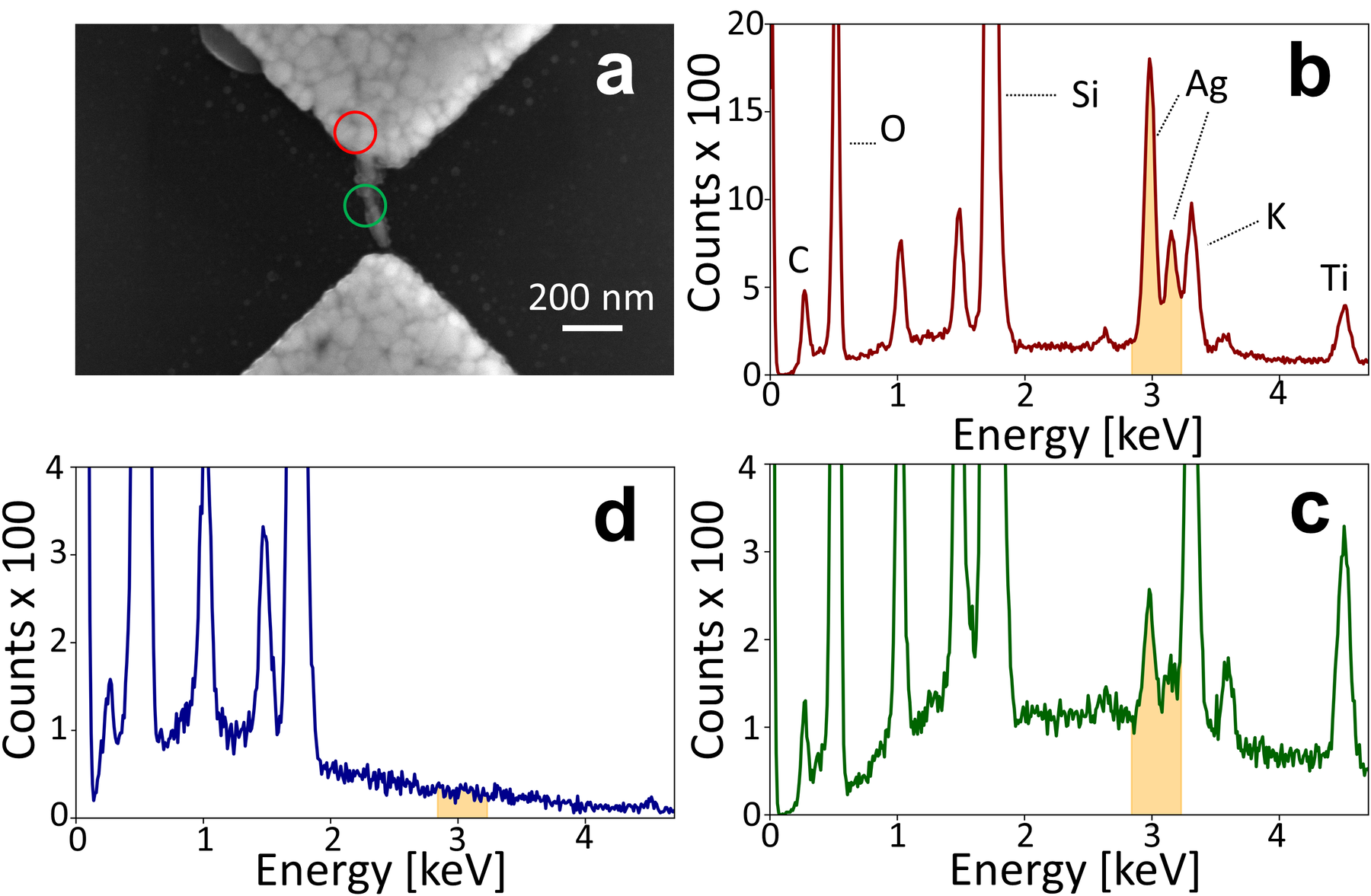}
\caption{\textbf{a} SEM image of a typical activated gap taken before the EDS measurements. The colored circles indicates where the EDS spectra were measured. \textbf{b}, \textbf{c} and \textbf{d} are EDS spectra corresponding to the silver electrode, the filamentary region, and the glass, respectively. The glass spectrum (d) was measured outside the area enclosed by the SEM image.}

 \label{fig:EDS}
\end{figure}

To confirm the nature of the material diffusing in the gap, we perform Energy-Dispersive X-ray Spectroscopy (EDS)~\cite{yang2018} on a set of devices bearing the same geometrical characteristics. Due to the fragile nature of filament formation and its rapid deterioration under the electron beam, we were unable to obtain a complete elemental map of the entire junction. Instead, we confirm the presence of silver in the gap region by performing point measurements at specified locations as shown in Figure~\ref{fig:EDS}. The main characteristic peak of silver corresponds to the Ag L$\alpha_1$ transition located at approximately 2.98 keV (Figure~\ref{fig:EDS}b). Higher-energy transitions, particularly the Ag L$\beta_1$ line at 3.1–3.3 keV, are more difficult to distinguish from other spectral features. In the spectrum measured on the filament between the two electrodes (Figure~\ref{fig:EDS}c), a silver peak is clearly observed at 2.98 keV, confirming the presence of Ag in this interelectrode region. The intensity of the Ag L$\alpha_1$ line measured on the filament remains lower than that on the electrode because of the smaller volume of silver atoms present in the gap compared to the silver electrode. The reference spectrum obtained from the glass substrate (Figure~\ref{fig:EDS}e) shows peaks associated with silicon and oxygen, as expected from a SiO$_2$ medium. The lack of a notable silver contribution in this spectrum confirms that the Ag lines observed in the gap region solely originates from the Ag material diffusing from the electrode during the activation sequence of the device and the growth of the filament. 


\section{Electroluminescence of the device}

\begin{figure}[htbp]
    \centering
        \includegraphics[width=\columnwidth]{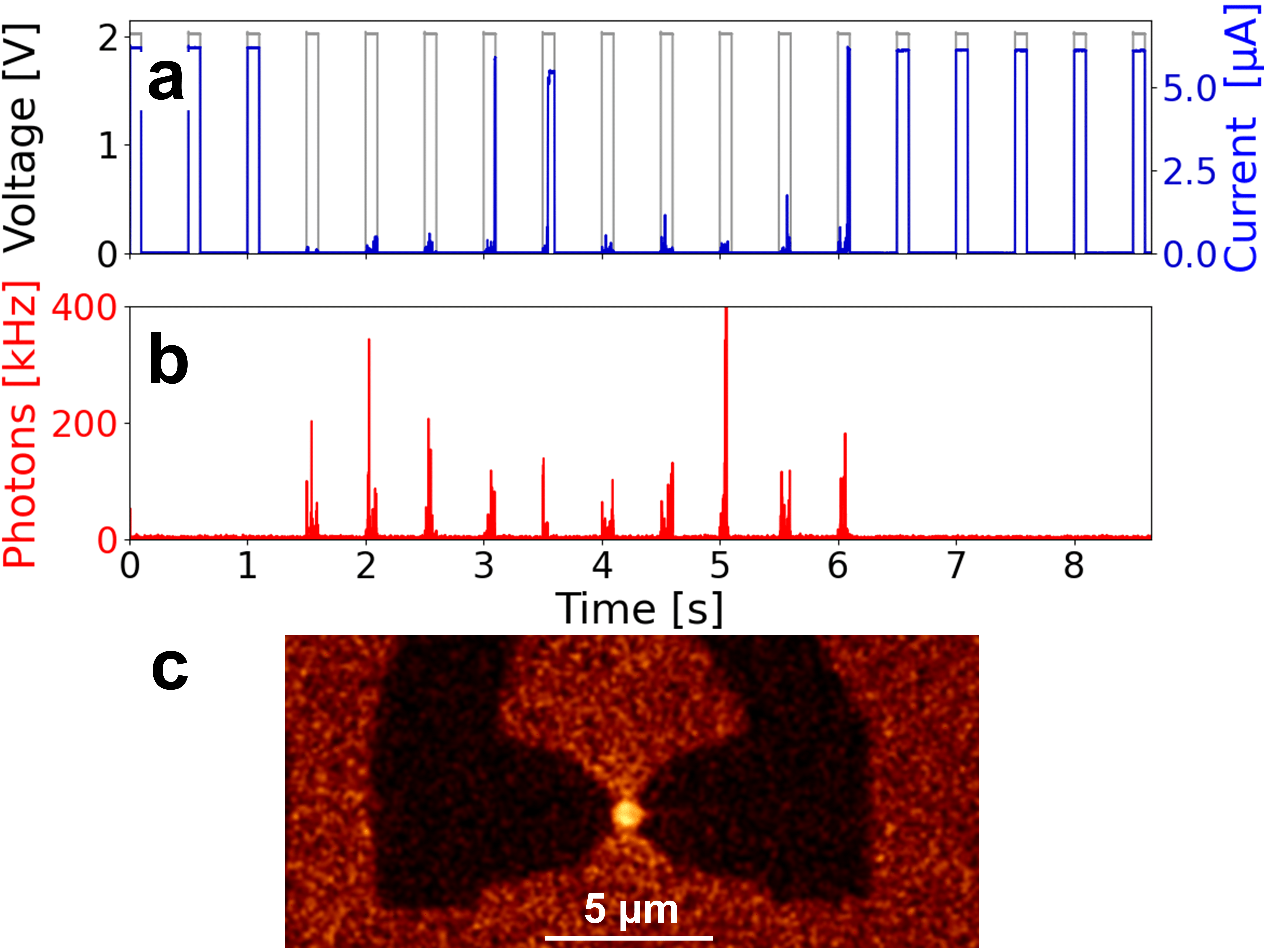}
    \caption{\textbf{a} Time plot of the voltage and current signals during a sequence of voltage pulses following the activation phase. $V=2$~V,  $T=500$~ms and $D=0.2$. \textbf{b} Corresponding electroluminescent trace. \textbf{c} Composite false-color CCD image of the memristor during operation. The overlay consists of an image acquired during the pulse sequence combined with an image of the sample observed under a weak illumination and in absence of an applied voltage. EL is restricted to the gap region. The electrodes are recognized as the darker grey areas.
 }
 \label{fig:EL}
\end{figure}

The formation and dynamical behavior of the Ag filament not only affect the memristor’s conductance but also play a crucial role in triggering an electroluminescent activity. This phenomenon is illustrated in Figures~\ref{fig:EL}a and \ref{fig:EL}b. During this sequence of voltage pulses ($V=2$~V, $T=500$~ms, $D=0.2$) the memristor’s conductance alternates between two states, reflecting the limited stability of the conductive bridge at that stage of the process. 
For the first three pulses, the current remains at a steady maximum, limited by the protection resistance here set at $R=325$~k$\Omega$. This compliance state, occurs when the resistance of the memristor becomes negligible compared to the protection resistance $R$. However, the initial filament connected the electrodes is not robust enough to sustain prolonged stimulation, and a spontaneous rupture of the connection after the third pulse is observed as a signifciantly reduce current level at the 4th pulse ($t$=1.5~s). Once broken, the conduction path restarts its growth dynamics with the next pulses as shown by the fluctuating current levels up until $t$=6~s. The current eventually reaches compliance again at the end of pulse 13. This newly formed pathaway is now stable enough to sustain prolonged stimulation at this voltage level, as no further changes in the device state is observed during the sequence.

Figure~\ref{fig:EL}b shows the simultaneously recorded EL counts detected using the avalanche photodiode. The device emits bursts of light exclusively during periods of current fluctuation – when the filament and its components are unstable. When the current is at compliance, the applied voltage is almost entirely dropped at the protection resistance  $R$ and EL disappears. This was systematically measured on all of the tested devices (12 memristive gaps). This observation aligns with previous studies on similar memristor architectures, where EL was linked to luminescent trapping sites, such as oxygen vacancies or silver aggregates, formed in the switching layer during filament growth~\cite{Cheng2022,Hamdad2023,Malchow2024}. The light emission is restricted to the gap region, as shown in Figure~\ref{fig:EL}c, which overlays a CCD image (integrated over the entire sequence) with a bright-field transmission image of the sample.  
It is the type of EL dynamics displayed in Figure~\ref{fig:EL}b that motivated us to investigate the early-stage formation of light-emitting defects within the dielectric matrix. 

Our measurement principle relies on detecting the photoluminescence (PL) of optically active species generated and diffusing in the gap during the forming sequence, anticipating the onset of switching – that is, before any current flows through the device.

\section{Photoluminescence response during the activation phase of the memristor}

The glass coverlips used as substrates and the PMMA matrix are known to generate a large photoluminescent background when illuminated by a focused green laser beam. To mitigate this contribution, we continuously scan under the focused laser spot a $25 \times 25$~$\mu$m area centered at the tested memristor with a moderated laser power $P=$~500 $\mu$W for an hour prior to the electrical activation. Hence, all unstable PL-emitting intrinsic defects at the glass interface as well as those present within the PMMA matrix are photobleached. The stable remaining PL is considered the background signal, which is typically reduced  by 10 fold compared to the PL level measured before the bleaching step. This prehemptive photobleaching ensures that any temporal changes observed on the PL maps during the activation only come from the device evolution triggered by the applied voltage pulses and not from any laser-induced dynamics of native emission centers present within the substrate. The experimental protocol discussed in the following section was applied to 20 devices. While there is some inherent gap-to-gap variability of the results, the conclusions we draw for the present case qualitatively applies to all tested devices. 

\begin{figure*}[!ht]
    \centering
        \includegraphics[width=2\columnwidth]{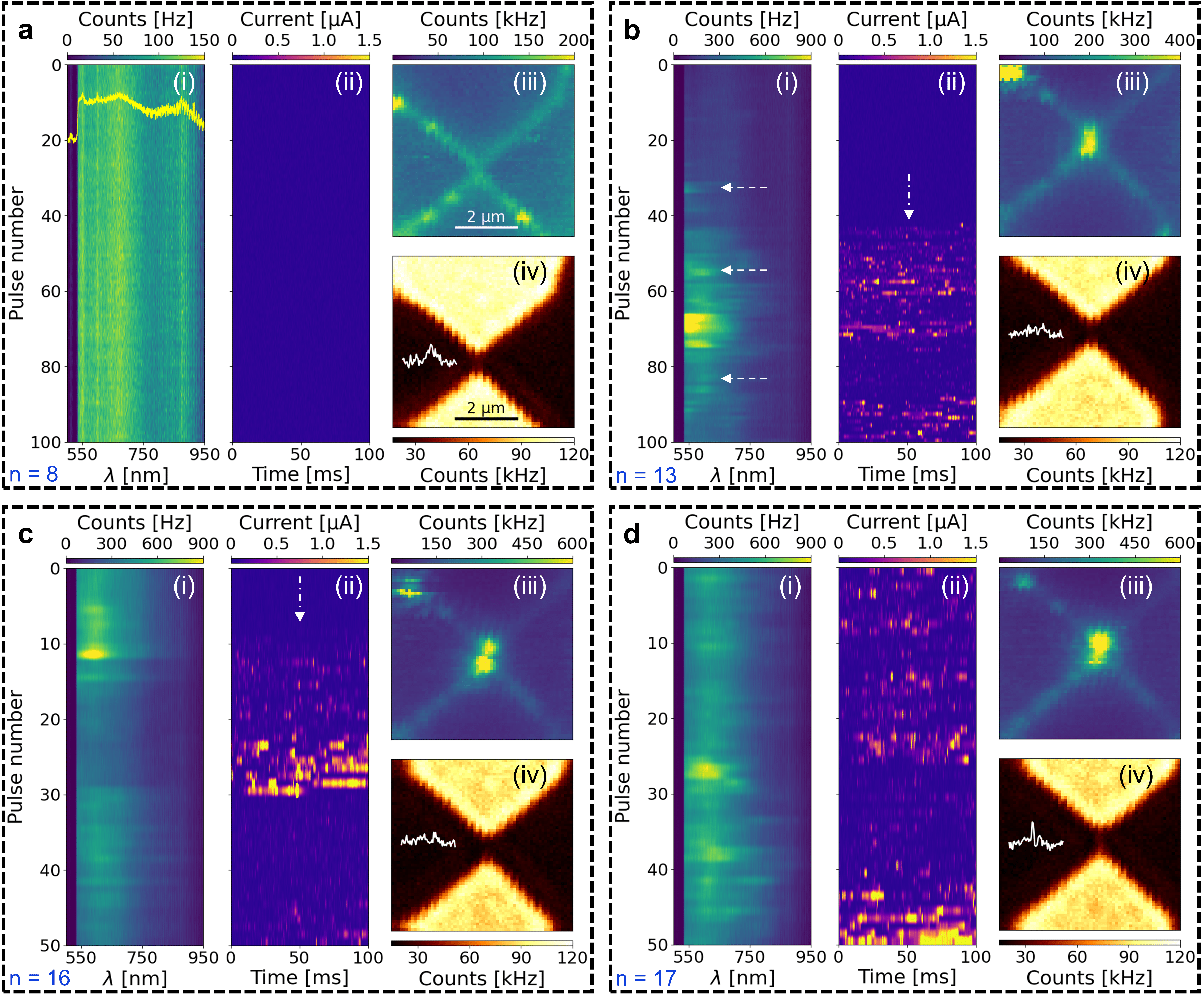}
\caption{\textbf{a} Reconstructed pulse-to-pulse evolution of the PL spectrum (i) and the current (ii) during the activation sequence $n=8$. The laser power is $P=50$~$\mu$W and the integration time is equal to the duration of the voltage pulse (100 ms). All spectra are corrected for the background noise of the CCD and the quantum efficiency of the detection path (transmission of the objective, CCD and grating efficiencies). The spectrum displayed in (i) is the average spectrum over the entire sequence. (iii) and (iv) are confocal scans of the PL and the laser reflection obtained after the pulse sequence. An image-wide profile across the gap is shown in white in (iv). \textbf{b}, \textbf{c} and \textbf{d} are similar measurements taken for sequences $n=13$, $n=16$, and $n=17$, respectively. Horizontal arrows in \textbf{b}(i) are pointing at various types of the spectral fluctuations such as intensity flare, spectral shift, and intermittency. The vertical arrows in \textbf{b}(ii) and \textbf{c}(ii) mark the onset of current flow during the sequence.}
\label{fig:dynamics}
\end{figure*}

The buildup of the conduction channel is achieved by applying $n$ activation sequences consisting of $N$ pulses with $V= 2\ \mathrm{V}$, $T = 500\ \mathrm{ms}$, $D = 0.2$. For the purpose of the experiment, the protection resistance $R$ is removed. Without compliance, the current buildup of the memristor is allowed to explore a larger parameter space. 

The PL activity of the memristor is monitored during the activation phase by two sequential protocols. During the $n$ sequences, we monitor the PL spectrum emitted by the gap at each voltage pulse with the help of the spectrometer. We set the integration time equal to the pulse period. Simultaneously, we record the current flowing in the device. All these spectrum-current pairs are then concatenated over the complete sequence of $N$-pulse to generate two complementary maps: spectral fluctuations of the PL and current variations. With this first set of measurements, we can follow the signal's dynamics during the sequence. Figures~\ref{fig:dynamics}a(i) and \ref{fig:dynamics}a(ii) display reconstructed maps of the PL spectra and the current for activation sequence $n=8$ (earlier sequences featured qualitatively and quantitatively similar signals) for a given memristor. The vertical axis represents the pulse number $N$. In Figure~\ref{fig:dynamics}a(i), the spectrum is steady across the $N$ pulses and reflects PL emitted by the silver, some remaining background signal from the glass as well as weak Raman lines of PMMA, probably enhanced by the presence of Ag. For this first sequence, no current is detected (see Figure~\ref{fig:dynamics}a(ii)).

Figures~\ref{fig:dynamics}a(iii) and \ref{fig:dynamics}a(iv) are post-sequence confocal images of the spatial distribution of the PL and the reflected laser signal obtained by scanning the sample through the laser focus. This set of data constitutes the second type of measurement. The PL map highlights locations of permanently generated optically-active defects, and the reflected laser intensity helps us to monitor structural changes. The PL image, Figure~\ref{fig:dynamics}a(iii), confirms the spectral map shown in Figure~\ref{fig:dynamics}a(i). The silver edges generate clear photoluminescence standing out from the remaining stable autofluorescence background of the glass coverslip. The reflection map in Figure~\ref{fig:dynamics}a(iv) unambiguously reveals the electrode terminals with the metal reflecting the 515 nm laser.

Figure~\ref{fig:dynamics}b shows the same set of measurements taken at $n=13$, which marks the first sequence where current flow was detected. At this stage, the PL spectra are no longer steady. Beyond $N=32$, the behavior is characterized by intense flares, spectral shifts, and intermittent activity (as indicated by the arrows in Figure~\ref{fig:dynamics}b(i)), all of which point to a significant increase in PL activity within the memristive gap. This dynamic behavior is further supported by the pulse-to-pulse current map displayed in Figure~\ref{fig:dynamics}b(ii), where a highly fluctuating current path emerges starting at $N=42$, an expected outcome given the absence of compliance to protect the memristor. In this case, Joule heating contributes to filament instability by initiating localized degradation of the conductive pathways~\cite{roldan2021}. While no clear correlation exists between the magnitude of the current fluctuations and the spectral changes, it is notable that variations in the PL spectra begin at $N=32$, preceding the emergence of measurable current.

The confocal PL and reflection images of the gap captured immediately after the voltage sequence are displayed in Figures~\ref{fig:dynamics}b(iii) and \ref{fig:dynamics}b(iv), respectively. The PL intensity at the gap is significantly enhanced, forming a symmetric two-lobe pattern aligned with the electrode tips, whereas the reflection map reveals no discernible changes. The increased PL signal in the gap was systematically observed across all devices, with enhancement factors ranging from 1.4 to 6.5. Because Ag diffusion is a field-assisted process~\cite{Hsiung2010}, the tapered electrode geometry ensures that filament growth initiates at the tips, thereby confining the spatial extent of the PL response to the gap. This experiment cannot definitively identify the species causing the enhanced photoluminescence at the gap, as several potential luminescent sources – including oxygen vacancies, Si-rich dots, and Ag clusters – emit within the same spectral range~\cite{Salh2011,Aceves14,Lee2004}. However, the SEM image of the filament (inset of Figure~\ref{fig:activation}) and the EDS measurements provide some insight. Given that the conduction pathways develop mainly at the interface, native defects burried in the SiO$_2$ are probably not involved. The PL observed in the gap is therefore most likely associated with Ag molecular aggregates, which form, migrate, and coalesce into a filament during the forming process~\cite{Wang2019,Hsiung2010}. At that point of the activation, however, the concentration of metal aggregates remains too small to induce any change in the laser reflectivity, despite the build-up of a conduction channel.

Applying repeated activation sequences injects more Ag clusters into the gap region. Figures~\ref{fig:dynamics}c and \ref{fig:dynamics}d correspond to sequences $n=16$, and $n=17$ of the activation process. In both cases, the PL spectrum remains unstable, exhibiting pulse-to-pulse variability. This behavior is again accompanied by a dynamic evolution of the current paths signalling the construction of conduction channels at every pulse. For $n=16$, the confocal images, Figure~\ref{fig:dynamics}c(iii) and Figure~\ref{fig:dynamics}c(iv), are qualitatively and quantitatively similar than those obtained after $n=13$. However, after $n=17$, the confocal PL image in Figure~\ref{fig:dynamics}d(iii) displays an asymmetric signature, with the highest intensity localized at the active electrode, which is the source of Ag ions. This asymmetry remains for subsequent sequences (data not shown). The particularity of the $n=17$ sequence is that a current is measured at the very first pulse ($N=1$) whereas for previous activation sequences, a few pulses where required to buildup the conduction channel (vertical arrows at $N=42$ in Figure~\ref{fig:dynamics}b(ii) and at $N=8$ in Figure~\ref{fig:dynamics}c(ii)). $n=17$ is therefore tagging the transition between volatile conduction channels, which are relaxing between sequences, to non-volatile current pathways. 
At this point of the activation, the filament becomes visible in the laser reflection image (Figure~\ref{fig:dynamics}d(iv)), and marks the effective completion of the activation. In the next sequence ($n=18$), we successfully recorded an electroluminescent response (not shown, but comparable to Figure~\ref{fig:EL}c.)

\begin{figure}[htbp]
    \centering
        \includegraphics[width=\columnwidth]{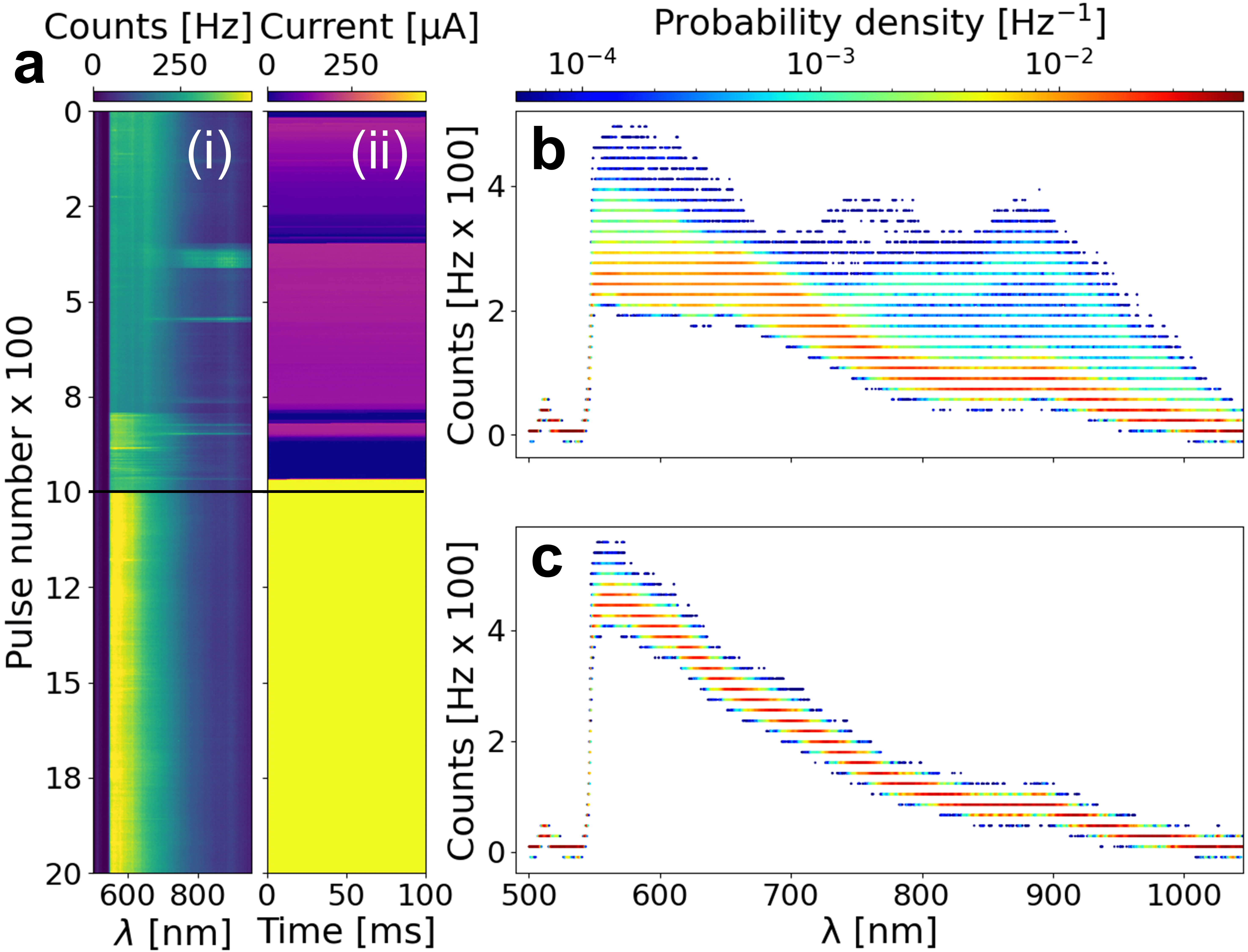}
\caption{\textbf{a} Pulse-to-pulse evolution of the PL (i) and current (ii) measured during a regime of unstable current level 
($N<980$) and while current is at compliance ($N>980)$. \textbf{b},\textbf{c} Probability densities of the spectral PL fluctuations measured under unstable current conditions, and at current compliance, respectively.}
\label{fig:compliance}
\end{figure}

Fluctuations in photoluminescence and the emergence of EL emission are both associated with irregular current levels, which point to atomic restructuring of the filament during the voltage sequence (see EL trend in Figure~\ref{fig:EL}c). Figure~\ref{fig:compliance}a shows PL and current maps obtained from another device and concatenating two consecutive voltage sequences (separated by the dashed horizontal lines) transitioning from an unstable current to a compliance level determined by a protection resistance ($R=6.12$~k$\Omega$). The transition takes place at pulse $N=980$. Clearly, the degree of PL fluctuations in these two regimes are different. To quantify this, we introduce the probability density shown in Figures~\ref{fig:compliance}b and \ref{fig:compliance}c~\cite{Margueritat2012}. When the current varies, we observe rare but significant changes in the PL spectrum (Figure~\ref{fig:compliance}b) compared to the compliance condition during which the PL spectrum stabilizes (Figure~\ref{fig:compliance}c). 

\begin{figure}[htbp]
    \centering
        \includegraphics[width=\columnwidth]{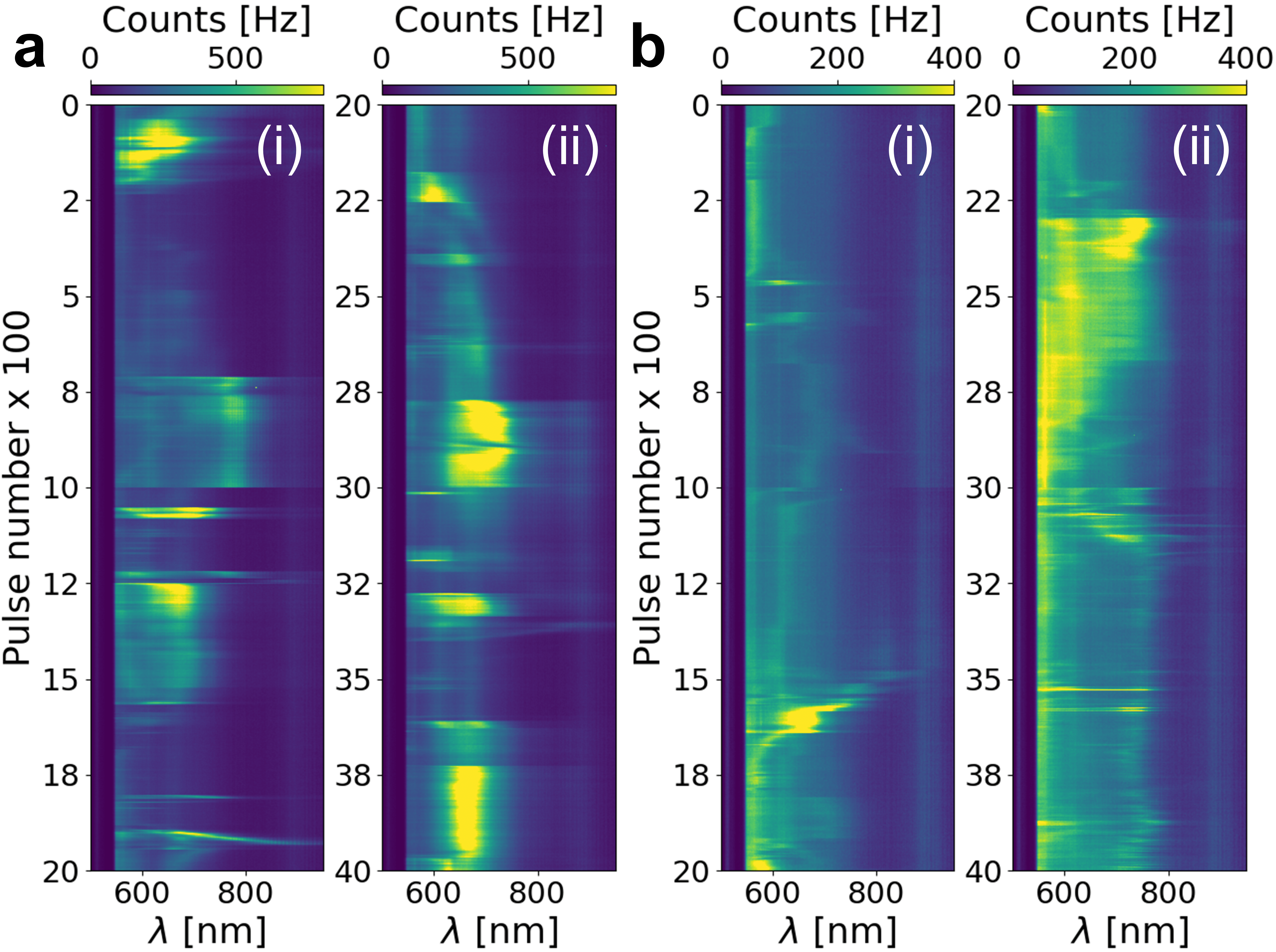}
    \caption{
    \textbf{a} Pulse-to-pulse evolution of the PL obtained by concatenating four activation sequences of 1000 pulses each, with
    $V=3$~V,  $T=500$~ms and $D=0.2$, all measured before the onset of current. Subfigure (i) shows pulses $0\leq N \leq 2000$, and (ii) shows pulses $2000 < N \leq 4000$.
    \textbf{b} Pulse-to-pulse evolution measured on a different device also before the detection of a current flow.}
    \label{fig:beforecurrent}
\end{figure}

Finally, as already alluded to in Figure~\ref{fig:dynamics}b, PL dynamics do not always coincide with current instabilities, but can even precede the onset of the current. Figure~\ref{fig:beforecurrent}a show reconstructed PL maps from yet another device for two consecutive activation sequences -- (i) and (ii) -- obtained before any current was recorded. The PL is a highly dynamic signal across the entire sequence with strong pulse-to-pulse variability such as large change of intensity (e.g. around $N=1100$ and $N=2800$), and band appearance or disappearance (e.g. $N=750$, $N=1000$), but also continuous intensity and spectral evolution (e.g. $N=1900$, $N>3800$). Figure~\ref{fig:beforecurrent}b is another example of the PL fluctuations obtained from a different memristor, also highlighting the drastic modification of the PL response during this early stage of the activation phase, resulting from the injection of Ag clusters. The PL spectrum most probably reflects the pulse-to-pulse variation of their size and concentration as well as their internal photodynamics -- such as brightness and bleaching.

\section{Conclusion}

This study provides a comprehensive investigation into the activation dynamics and light-emission mechanisms of Ag-based planar memristors. By combining electrical excitation with \emph{in-situ} photoluminescence imaging, we reveal that the activation phase of the memristor introduces significant PL fluctuations within the memristive gap. These fluctuations originate from the generation, diffusion, and aggregation of Ag clusters that progressively form the conductive filament responsible for resistive switching. We find that the PL activity appears before any measurable current, highlighting the contribution of early-stage ionic and structural dynamics. 
Once resistive switching occurs, PL evolves from a transient phase to a stable regime, reflecting therefore the progressive evolution of the memristor’s internal architecture - from dispersed luminescent precursors to a semi-continuous metallic filament. We emphasize that PL measurements average the optical response over the entire gap region due to the diffraction-limited spatial resolution. Consequently, the spectral changes observed during activation and subsequent current fluctuations reflect the collective behavior of multiple emitting species within the gap. In contrast, EL is not constrained by diffraction-limited excitation and originates from electron injection into localized defects, potentially at the single-emitter level. Therefore, EL is directly coupled to charge transport pathways—requiring current flow—and to their dynamic reconfiguration, whereas PL captures precursor processes occurring prior to any measurable conduction. Despite this distinction, the combination of spatially averaged PL and temporally integrated EL provides a coherent picture: PL probes the early formation and evolution of emissive species, while EL reveals their role in electrically active conduction pathways at later stages. Both signals stem from the same activation process that generates emissive centers, although it remains uncertain whether identical entities are responsible for both emissions.

Overall, these results provide direct experimental evidence that light-emitting behavior in Ag-based memristors arises from nanoscale diffusive processes inherent to filament formation. This insight bridges the gap between electrical and optical functionalities in memristive systems and establishes a framework for engineering hybrid optoelectronic devices where electrical control and light emission coexist within a single nanoscale platform.

\medskip
\textbf{Data Availability Statement} \par 
The data that support the findings of this study are available from the corresponding author, Alexandre Bouhelier, upon reasonable request.

\medskip
\textbf{Acknowledgements} \par 
We thank Erik Dujardin and Phong Nguyen for fruitfull discussions. This work has been partially funded by the Région Bourgogne Franche-Comté throught theh ATRACT program, the European Regional Development Fund
(FEDER-FSE Bourgogne Franche-Comté 2021/2027), the French Agence
Nationale de la Recherche (ANR-25-CE24-7328 POMMARD), the EIPHI Graduate School (ANR-17-EURE-0002), and Nicolaus Copernicus University Center of Excellence (03.01.00000810). Device fabrication and characterization
were performed at the technological platforms ARCEN Carnot
and SMARTLIGHT with the support of the French Agence
Nationale de la Recherche under program Investment for the
Future (ANR-21-ESRE-0040), the
CNRS, and the French RENATECH+ network.

\medskip
\textbf{Conflict of Interest} \par
The authors declare no conflict of interest.
\pagestyle{plain}      

\bibliographystyle{MSP}
\bibliography{references}



\end{document}